\documentclass[twocolumn]{openjournal}
\pdfoutput=1

\usepackage{amsmath}
\usepackage{xcolor}
\usepackage{natbib}
\usepackage{needspace}

\begin{document}

\shorttitle{Effect of non-spherical projectiles on dust aggregates}
\shortauthors{Kolanz L., Lazzati D.}

\title{Effect of non-spherical projectiles on the structure of porous dust aggregates formed by coagulation}

\author{Lucas Kolanz, Davide Lazzati}

\affil{Department of Physics, Oregon State University, 301
  Weniger Hall, Corvallis, OR 97331, USA} 

%%%%%%%%%%%%%%%%%%%%%%%%%%%%%%%%%%%%%%%%%%%%%%%%%%
\begin{abstract}
Cosmic dust is ubiquitous in the universe, yet the structure and geometry of individual grains remain poorly understood. The existence of non-spherical, fluffy, and even fractal grain structures is predicted by numerical simulation and supported by observations of linear polarization of starlight. However, it has proven challenging to go beyond a qualitative investigation of such crucial grain characteristics. We present soft-sphere discrete element simulations of dust coagulation with sequential collisions using non-spherical projectiles of various sizes. We study the internal structure and geometry of the resulting aggregates under three growth conditions: a constant final aggregate size, a constant number of projectiles, and a constant projectile size. In most cases we allow for projectile internal restructuring, but we also test the effect of enforcing a constant projectile structure after impact. We find that aggregates' porosity and fractal dimension depend both on the size and number of projectiles, and that their asymmetry and stretch parameters depend more on the number of projectiles than on projectile size. Overall, the grain porosity increases with both the number of projectiles and the size of the individual projectiles. Comparison with constraints from interstellar polarization indicates that none of our sufficiently large aggregates have structures capable of reproducing the observed polarization of starlight in the interstellar medium. We conclude that cosmic dust undergoes additional processing after coagulation to acquire structures consistent with observations.
\end{abstract}

%%%%%%%%%%%%%%%%%%%%%%%%%%%%%%%%%%%%%%%%%%%%%%%%%%
\section{Introduction} \label{sec:intro}

Cosmic dust is an important but often overlooked component of our universe. Not only is it a fundamental building block of asteroids, comets, and planetesimals, but it also plays key roles in many physical processes. The presence or absence of dust can even change the outcome of star formation and supernova explosions \citep{Dopcke2011,Gonzalez2025} even though it only makes up a small fraction of these systems. 
Thus, understanding where it comes from and how it interacts with its environment is important for understanding our universe and how it evolves. How dust interacts with its environment, whether that be through radiation or other dust particles, depends in part on its structure and other physical properties. 

The structure of cosmic dust has been estimated using numerical methods such as N-body simulations \citep{Wada2007}, analytical methods such as EMT-Mie theory and the Discrete Dipole Approximation \citep{Ysard2018}, experiments \citep{Wurm2000}, and dust collected from the deep sea, space, and the stratosphere \citep{Brownlee1985}. 

Dust in the interstellar medium (ISM) also polarizes starlight and produces polarized thermal emission. Dust grains tend to rotate with preferred orientations relative to external magnetic fields. This alignment produces an orientation-dependent extinction cross section that polarizes starlight.

Recently, \citet{Draine2024b} put constraints on the structure of dust in the interstellar medium (ISM) using starlight polarization and polarized thermal emission. They found highly porous dust must be flattened/elongated sufficiently to explain this polarization. 

In \citet{Kolanz2026}, we used our soft-sphere discrete element method (ssDEM) code DECCO (Discrete Element Cosmic COllision) to simulate dust coagulation and explore the role of temperature on the structure of dust aggregates. In these simulations we grew aggregates one spherical monomer at a time, fully simulating each growth step to allow for the possibility of aggregate restructuring. 

In this paper, we again fully simulate the collision of each projectile with the growing target aggregate, but now projectiles are themselves aggregates of a set number of monomers. We then use the structural metrics explored in \citet{Kolanz2026,Draine2024a,Draine2024b} to quantify the effects of projectile size and the number of projectiles on the resulting aggregate structure. Additionally, we test the difference between treating projectile aggregates as groups of monomers bound together purely with attractive forces versus treating them as rigid bodies, essentially treating projectiles as non-spherical elements. 

Using non-spherical elements in DEM simulations has historically run into limitations. The main problems stem from efficient collision detection, which is greatly complicated with non-spherical elements. Still, non-spherical elements are necessary in some systems in order to reproduce the correct bulk behavior. Non-spherical elements are important, e.g., for studying such systems as hopper discharge rates \citep{Tangri2019}, mixing of different granular materials \citep{Govender2023}, and ballast mechanics \citep{Hou2023}.

This manuscript is organized as follows: in section \ref{sec:methods} we describe our simulation code and physical parameters, the aggregate growth schemes, and metrics for quantifying aggregate structure. In section \ref{sec:results} we describe the results of our simulations. Section \ref{sec:summary} summarizes and discusses our results and their implications. 

Throughout this work, the term ``particle'' refers to either a solid piece of dust or a dust aggregate. We use the term ``monomer'' to refer to an individual, spherical, solid dust particle. A monomer is the smallest unit treated in our simulations and has a radius of the order $\sim 0.1\,\mu\mathrm{m}$. The term ``aggregate'' denotes a bound collection of monomers held together by attractive forces or treated as a rigid body. A ``fragment'' is a projectile which is itself an aggregate, used to grow a larger target aggregate. We refer to the number of monomers in an aggregate or fragment as its ``size''. During a collision, the larger-sized grain is referred to as the target and the smaller as the projectile. If they are the same size, the target and projectile aggregates will be defined as needed. We group metrics of aggregate structure into two categories: the ``internal structure'', which refers to the porosity and fractal dimension, and the ``geometry'', which refers to the asymmetry and stretch parameters (defined in Section~\ref{subsec:structureQuantification}). In this study we focus on fragments as projectiles. However, in limiting cases, we include monomers as projectiles. 

\begin{figure*}
    \centering
    \includegraphics[width=\textwidth,height=0.93\textheight,keepaspectratio]{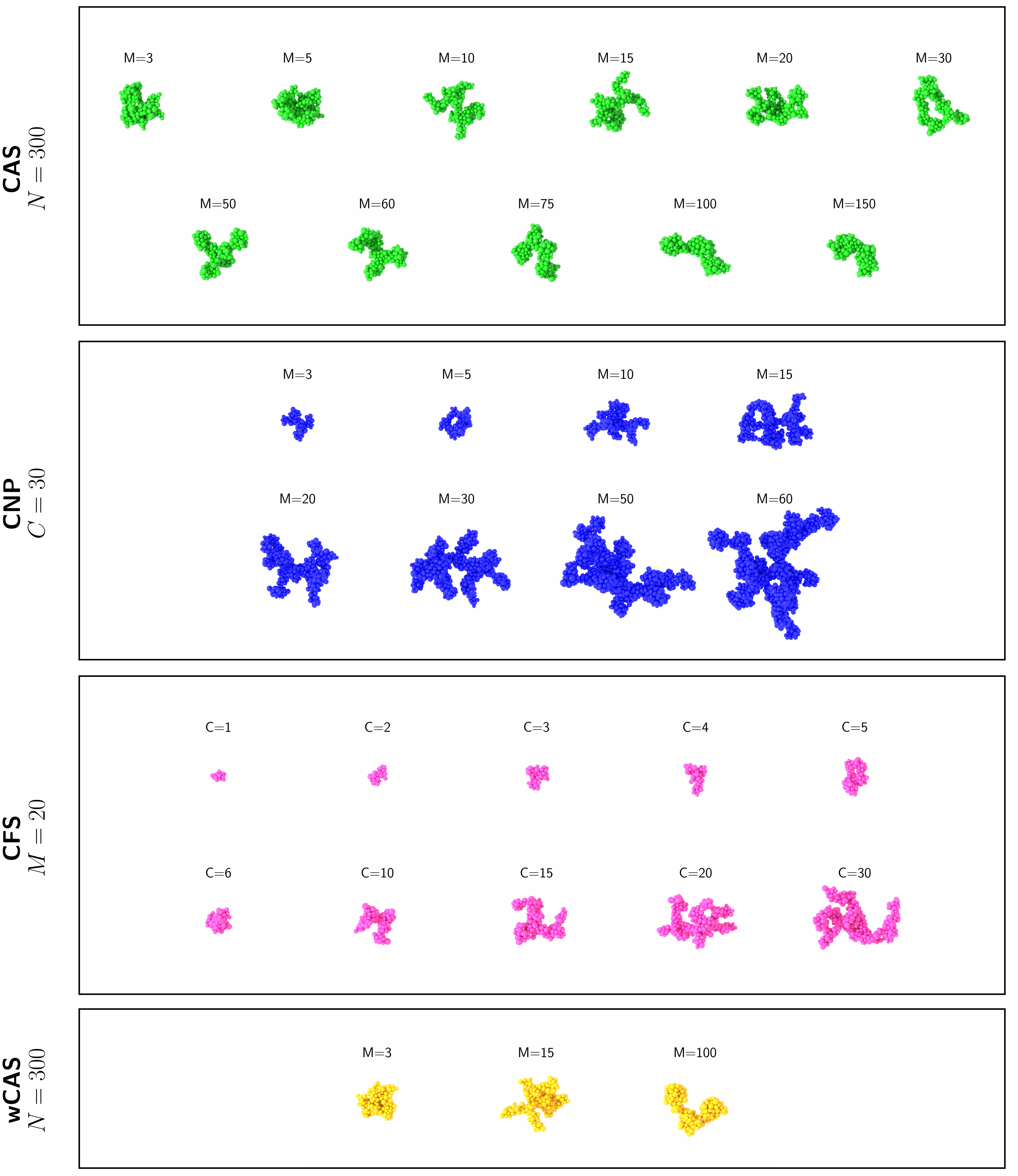}

    \caption{Example grains for all growth methods tested in this study. Growth methods are color-coded. Grains grown to a final size of $N=300$ monomers (CAS) are green, grains grown with $C=30$ fragments each (CNP) are blue, grains grown with fragments of size $M=20$ (CFS) are pink, and grains grown to a final size of 300 monomers using rigid body fragments (wCAS) are yellow. The fragment size $M$ (or number of projectiles $C$) used is listed above each grain.} 
        \label{fig:aggStructure}
\end{figure*}

\section{Methods}\label{sec:methods}

\subsection{Numerical Simulation}
\label{subsec:NumSim}

For this study we use the DECCO code \citep{Guidos2025,Kolanz2026} to run three-dimensional, ssDEM dust coagulation simulations to produce dust aggregates. Projectiles in this study are also aggregates of various sizes grown in \citet{Kolanz2026}. These aggregates are made up of monomers whose radii were drawn from a lognormal distribution. The standard deviation of the lognormal distribution is set to $\sigma=0.2$. Monomers in the final aggregates have radii ranging from 0.04 to 0.20\,$\mu\mathrm{m}$. The final aggregate radii range from 0.29 to 4.68\,$\mu\mathrm{m}$, measured as the distance from the aggregate's center of mass to the furthest monomer's center.

Monomers have a density of $\rho=2.25$\,g/cm$^3$, consistent with graphite. Depending on external conditions, the coefficient of friction of graphite can vary greatly. We use a value of $\mu_k$=0.1 \citep{morstein2022} for the coefficient of sliding friction and $\mu_r=10^{-5}$ for the coefficient of rolling friction, both at the low end. These are the same parameters used in \citet{Kolanz2026}.

DECCO treats contact forces between monomers as linear springs with a constant coefficient of restitution. If $k_{in}$ and $k_{out}$ are the spring constants during compression and decompression, respectively, the coefficient of restitution is calculated as 

\begin{equation}
    \epsilon=\sqrt{k_{out}/k_{in}}.
    \label{eq:cor}
\end{equation}

Since $k_{in} > k_{out}$, this saps energy from the system, mimicking energy dissipation from plastic deformation. It is also applicable for particles of various sizes.
In order to treat the soft, compressible nature of real matter, $k_{in}$ is set such that there will be multiple time steps calculated for any collision in the simulation. Then $k_{out}$ is set based on equation~\ref{eq:cor}.

The code used here is the same as that used in \citet{Kolanz2026}, with the addition of rigid body physics. Aggregates treated as rigid bodies consist of monomers that cannot move relative to one another. This is known as the clumped or multi-sphere method \citep{Favier1999}. We follow \citet{Favier1999} for our implementation. However, we use quaternions to track rotations and move between reference frames \citep{Kartiwa2023}. 

To calculate forces on a rigid body, monomer-monomer forces and torques are calculated for all pairs of monomers not part of the same rigid body\footnote{Quantities labeled with the ``body'' subscript are in the body-fixed frame and those without are in the inertial simulation frame.}. The total force on a rigid body is then calculated as

\begin{equation}
    \vec{F}_{j} = \sum_{i\in G_{j}} \vec{F}_{i},
\end{equation}

\noindent where $\vec{F}_{j}$ is the total force on body $j$, $\vec{F}_{i}$ is the total force on monomer $i$ from all other monomers not in group $j$, and $G_j$ is the set of monomers in body $j$. The body's acceleration is then simply

\begin{equation}
    \vec{a}_{j}=\vec{F}_{j}/m_j,
\end{equation}

\noindent where $m_j$ is the mass of body $j$.

The total torque on a rigid body is calculated as

\begin{equation}
    \vec{\tau}_{j}=\sum_{i\in G_{j}} (\vec{r}_{i,j} \times \vec{F}_{i} + \vec{\tau}_{i})
\end{equation}

\noindent where $\vec{\tau}_{j}$ is the total torque on body $j$, $\vec{r}_{i,j}$ is the displacement vector from the center of mass of body $j$ to the center of monomer $i$, and $\vec{\tau}_{i}$ is the total monomer-monomer torque experienced by monomer $i$. 

The body's angular acceleration is then calculated as 

\begin{equation}
    \vec{\alpha}_{body,j} = \boldsymbol{I}_{body,j}^{-1}(\vec{\tau}_{body,j} -\vec{\omega}_{body,j}\times \boldsymbol{I}_{body,j}\vec{\omega}_{body,j}) 
\end{equation}

\noindent where $\vec{\alpha}_{body,j}$ is the angular acceleration, $\boldsymbol{I}_{body,j}$ is the moment of inertia tensor, $\vec{\tau}_{body,j}$ is the total torque, and $\vec{\omega}_{body,j}$ is the angular velocity, all of which are of body $j$ and in the body-fixed frame. 

A body's velocity $\vec{v}_{j}$, angular velocity $\vec{\omega}_{body,j}$, and center of mass position $\vec{r}_{j}$ are then found by integrating with the half-step Verlet method \citep{Verlet1967}. 

Individual monomer velocities and angular velocities are then obtained using

\begin{equation}
    \vec{v}_i = \vec{v}_{j} + \vec{\omega}_{j} \times \vec{r}_{i,j}
\end{equation}

\begin{equation}
    \vec{\omega}_i = \vec{\omega}_{j}
\end{equation}

\noindent where $\vec{v}_i$ is the velocity of monomer $i$, $\vec{v}_j$ is the velocity of body $j$, and $\vec{\omega}_i$ is the angular velocity of monomer $i$.

Translational quantities are stored in the inertial simulation frame, while angular quantities are stored in the body-fixed frame to avoid recalculating the moment of inertia tensor in the inertial frame for each body at each timestep.

We rotate between frames as needed using quaternions. A vector quantity of body $j$ (for example, $\vec{p}_j$) can be rotated from the body-fixed frame to the inertial frame with

\begin{equation}
    \vec{p}_j = \vec{p}_{body,j} + 2q_{0,j}(\boldsymbol{q}_{v,j}\times \vec{p}_{body,j}) +2\boldsymbol{q}_{v,j}\times(\boldsymbol{q}_{v,j}\times \vec{p}_{body,j})
\end{equation}

\noindent where $\boldsymbol{q_j}$ is the unit quaternion describing the rotation from body-fixed to inertial frames for group $j$ with scalar component $q_{0,j}$ and vector component $\boldsymbol{q}_{v,j}$, and $\vec{p}_j$ and $\vec{p}_{body,j}$ are the representations of the vector in the inertial and body-fixed frames, respectively. 

The rotation from the inertial frame to the body-fixed frame is then

\begin{equation}
    \vec{p}_{body,j} = \vec{p}_{j} - 2q_{0,j}(\boldsymbol{q}_{v,j}\times \vec{p}_{j}) +2\boldsymbol{q}_{v,j}\times(\boldsymbol{q}_{v,j}\times \vec{p}_{j}).
\end{equation}

At each timestep $\boldsymbol{q}_{j}$ is updated with

\begin{equation}
    \boldsymbol{q}_{n+1,j}
    =
    \boldsymbol{q}_{n,j}
    \otimes
    \exp\left[
        \frac{dt}{2}
        \left(
            0,\vec{\omega}_{body,j}^{\,n+1/2}
        \right)
    \right]
\end{equation}

\noindent where $\boldsymbol{q}_{n,j}$ is body $j$'s quaternion at timestep $n$, $dt$ is the timestep, $\otimes$ is the quaternion product, and $\vec{\omega}^{n+1/2}_{body,j}$ is the half-step angular velocity of group $j$ in the body-fixed frame \citep{Kartiwa2023}.

\subsection{Aggregate Growth}
\label{subsec:aggGrowth}

\begin{figure*}
    \centering
    \includegraphics[width=\textwidth,keepaspectratio]{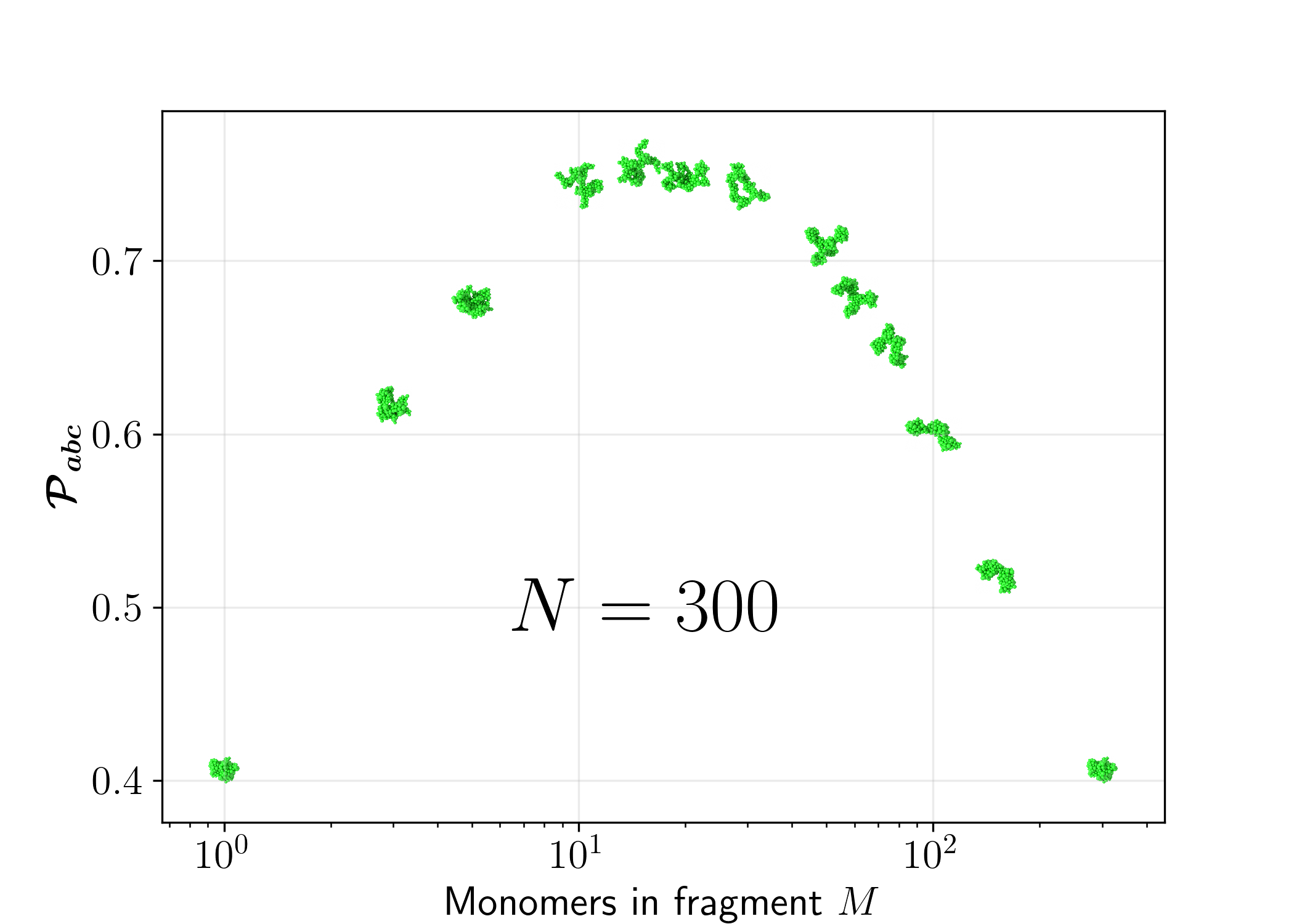}

    \caption{Equivalent ellipsoid porosity $\mathcal{P}_{abc}$ vs number of monomers in projectile fragments, $M$. Points are plotted with an image of a representative final aggregate to visualize their structural range. All aggregates in this figure are CAS aggregates, and consist of $N=300$ total monomers.} 
        \label{fig:CASPabcVsM}
\end{figure*}

\begin{figure*}
    \centering
    \includegraphics[width=\textwidth,keepaspectratio]{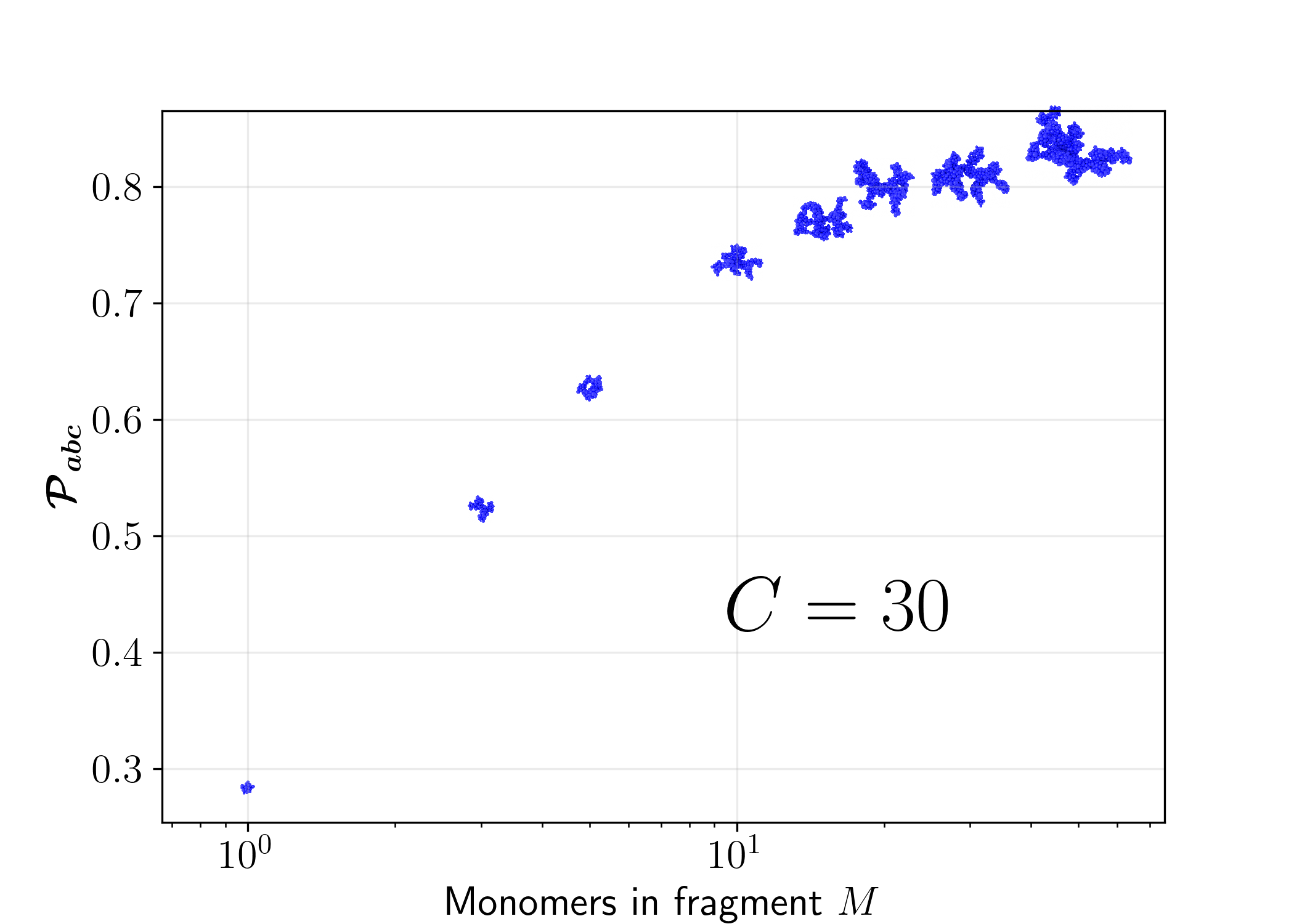}

    \caption{Equivalent ellipsoid porosity $\mathcal{P}_{abc}$ vs number of monomers in projectile fragments, $M$. Points are plotted with an image of a representative final aggregate to visualize the structural range. All aggregates in this figure are CNP aggregates, and were grown with $C=30$ projectiles each.} 
        \label{fig:CNPPabcVsM}
\end{figure*}

\begin{figure*}
    \centering
    \includegraphics[width=\textwidth,keepaspectratio]{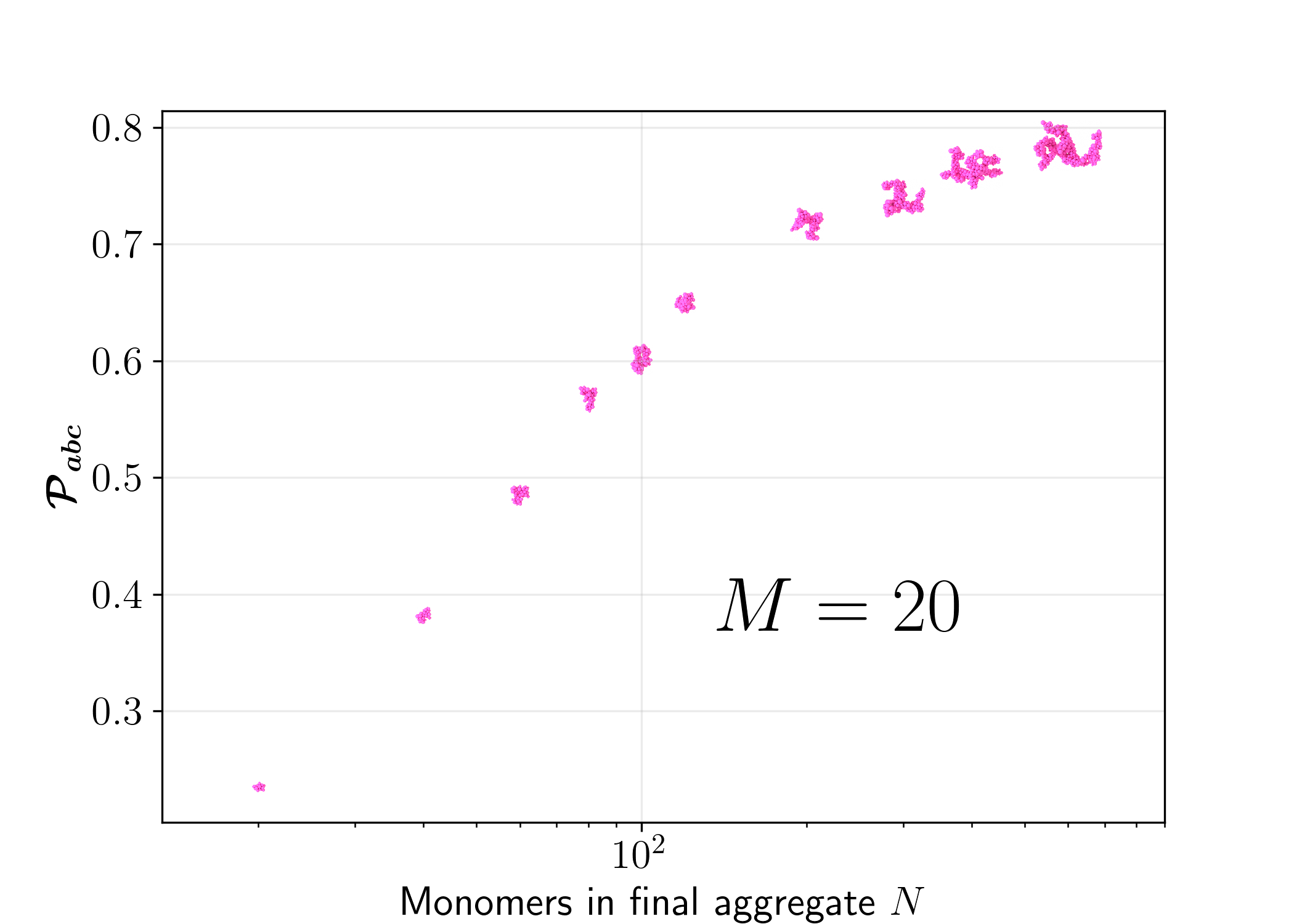}

    \caption{Equivalent ellipsoid porosity $\mathcal{P}_{abc}$ vs number of monomers in the final aggregate, $N$. Points are plotted with an image of a representative final aggregate to visualize the structural range. All aggregates in this figure are CFS aggregates, and were grown with a constant fragment size of $M=20$ monomers.} 
        \label{fig:PabcVsN}
\end{figure*}

\begin{figure*}
    \centering
    \includegraphics[width=\textwidth,height=0.93\textheight,keepaspectratio]{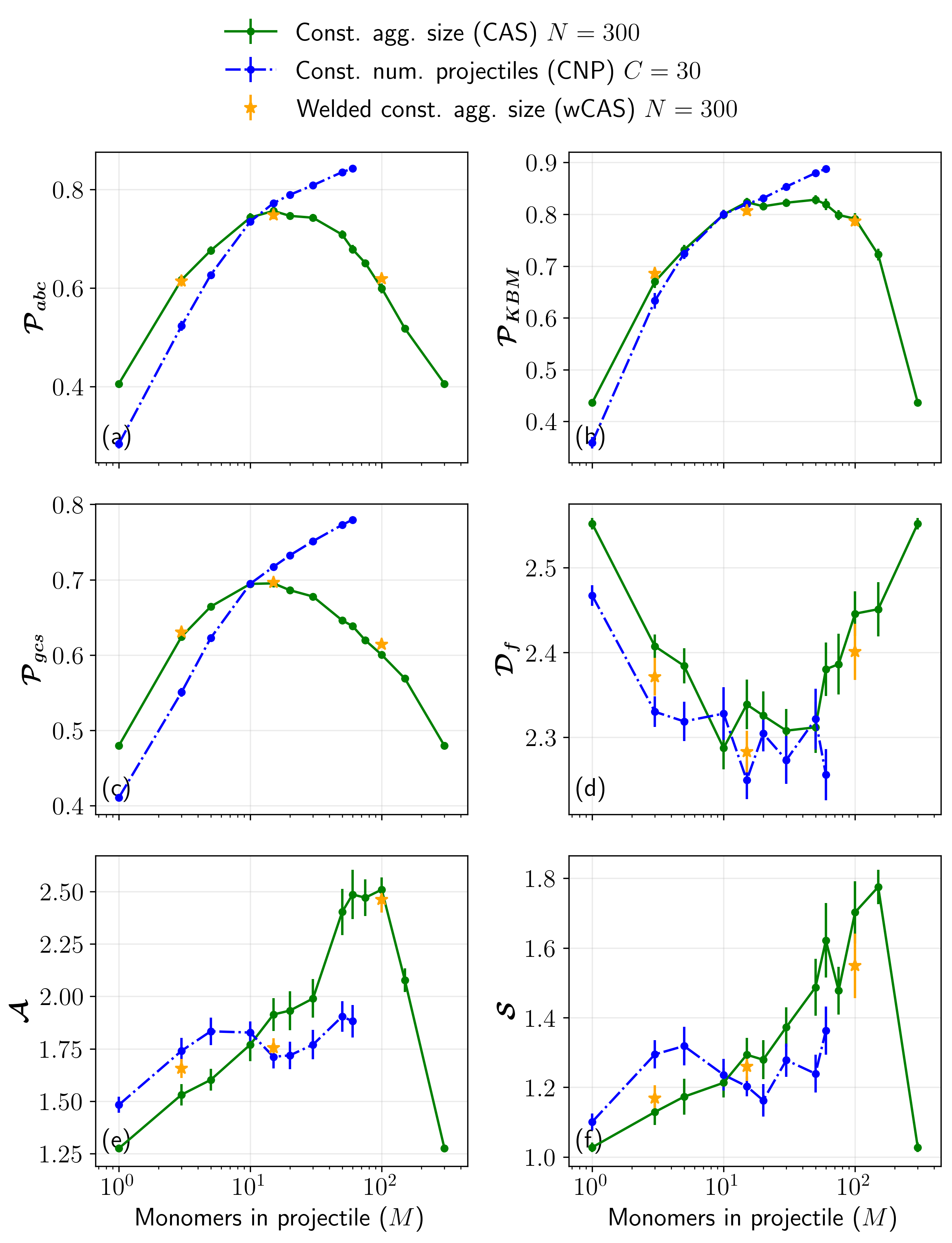}

    \caption{Structural metrics vs fragment size $M$ for CAS, CNP, and wCAS aggregates. The green line represents aggregates grown to a constant final aggregate size of $N=300$ (CAS). The blue line represents aggregates grown with a constant number of projectiles $C=30$ (CNP). The yellow points represent aggregates grown to a final size of $N=300$ where each projectile is a rigid body (wCAS).} 
        \label{fig:structural_metrics_vs_M}
\end{figure*}

\begin{figure*}
    \centering
    \includegraphics[width=\textwidth,height=0.93\textheight,keepaspectratio]{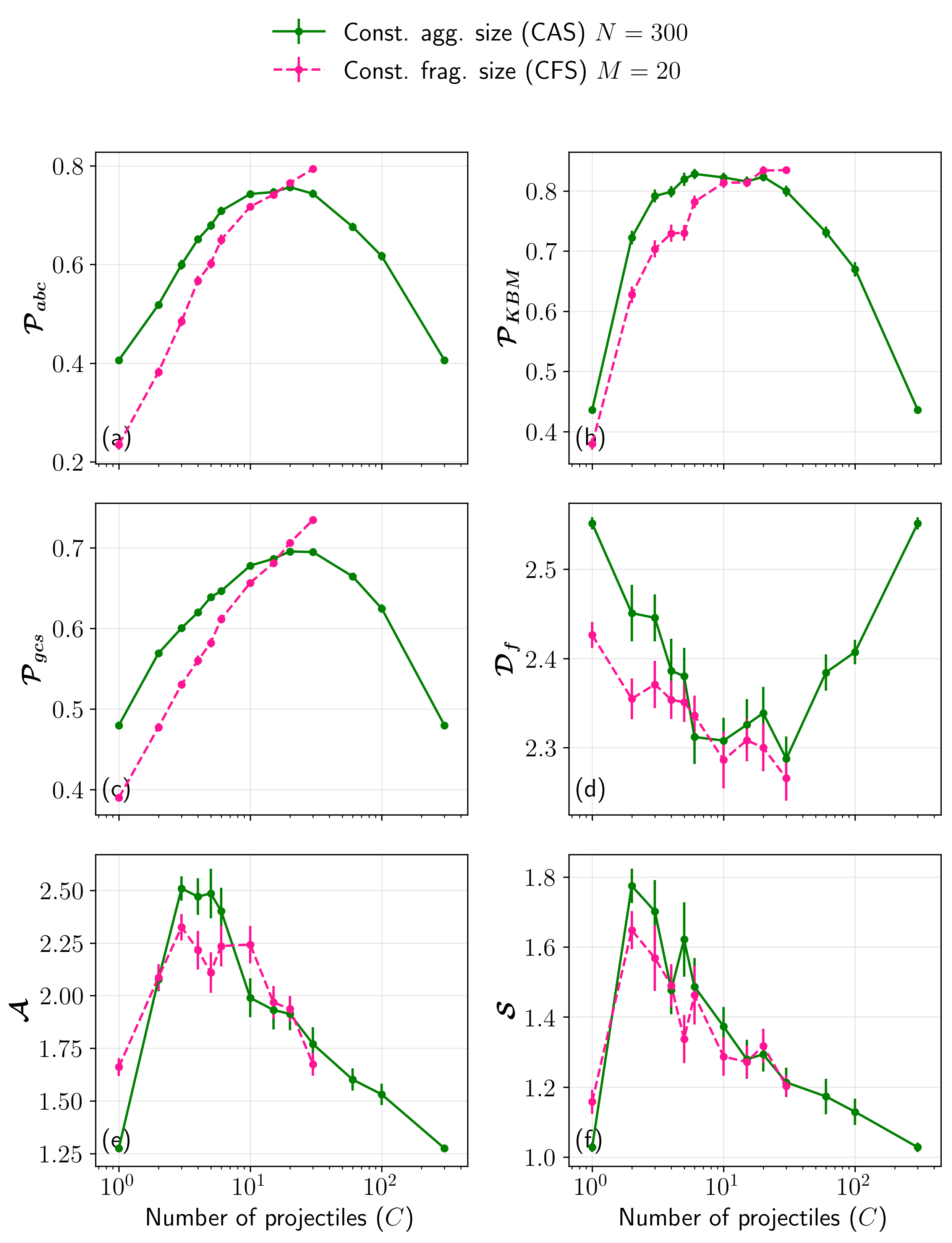}

    \caption{Structural metrics vs number of projectiles $C$ for CAS and CFS aggregates. The green line represents aggregates grown to a constant final aggregate size of $N=300$ (CAS). The pink line represents aggregates grown with a constant fragment size $M=20$ (CFS).} 
        \label{fig:structural_metrics_vs_C}
\end{figure*}

\begin{figure*}
    \centering
    \includegraphics[width=\textwidth,height=0.93\textheight,keepaspectratio]{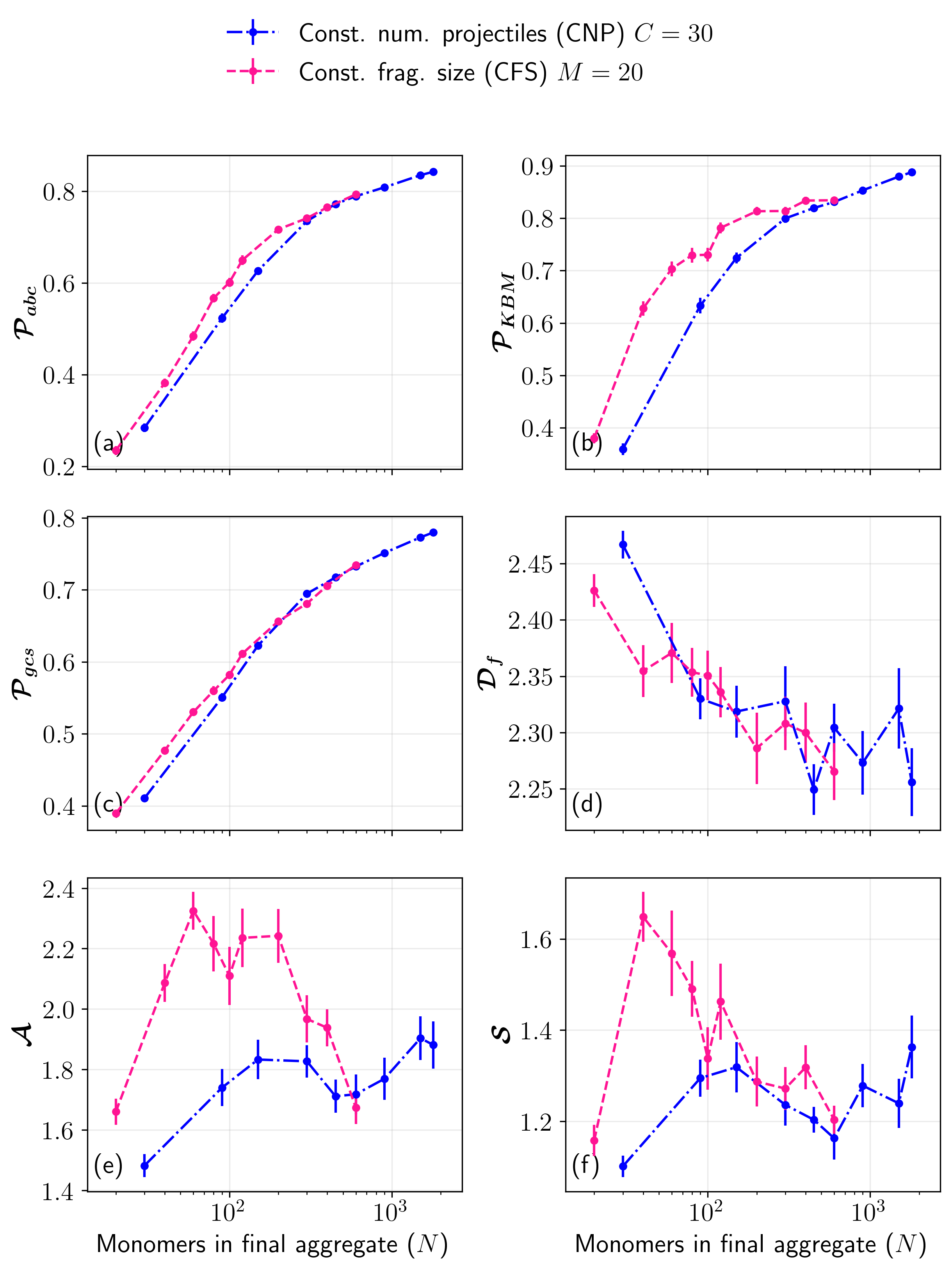}

    \caption{Structural metrics vs final aggregate size $N$ for CNP and CFS aggregates. The blue line represents aggregates grown with a constant number of projectiles of $C=30$ (CNP). The pink line represents aggregates grown with a constant fragment size of $M=20$ (CFS).} 
        \label{fig:structural_metrics_vs_N}
\end{figure*}

Aggregates in this study were grown using sequential collisions \citep{Suyama2008,Kolanz2026}. In sequential collisions, all growth steps are fully simulated, allowing for the possibility of aggregate restructuring. This differs from sequential sticking where projectiles are simply stuck on the growing aggregate and no restructuring is possible. In \citet{Kolanz2026} we grew aggregates one monomer at a time with sequential collisions and sequential sticking in order to compare them. We include the same sequential sticking data in some figures in this paper for the same reason.

We assume that the projectiles are in thermal equilibrium\footnote{The assumption of thermal equilibrium was used in \citet{Kolanz2026} to isolate and study the effect of thermal velocities on the structure of growing aggregates, and is kept in this study to stay consistent. In some astrophysical environments grains reach higher speeds due to magnetohydrodynamic turbulence \citep{Yan2004}.} with the surrounding gas. Their initial velocities are therefore drawn from a Maxwell-Boltzmann distribution determined by the projectile mass and ambient temperature and are thus strictly thermal. In general, aggregates form denser structures at higher temperatures because monomers have enough energy to restructure upon impacts but not enough energy to fracture the grain \citep{Kolanz2026}. A value of 1000 K is used as the temperature for all simulations in this paper. 

Projectiles in this study are themselves aggregates (which we term ``fragments'') which consist of various numbers of monomers (which we refer to as the aggregates' ``size''). These fragments are taken from a random aggregate at a given stage in its growth from \citet{Kolanz2026}. All aggregates are grown with fragments of a constant number of monomers. In \citet{Kolanz2026} aggregates naturally gain and lose angular momentum as additional monomers deposit onto them. Thus, each fragment's initial angular velocity is carried over from its last growth step. Fragments are given a random direction and offset before a collision. The offset is uniformly chosen in both dimensions of the plane perpendicular to the projectile's direction, and can range between plus or minus the target aggregate’s initial radius, calculated as the distance from the furthest monomer's center to the center of mass of the aggregate. Typical final aggregates for all growth schemes are shown in Figure~\ref{fig:aggStructure}.

The aggregates grown for this study can be parameterized by three values: the total number of monomers in the aggregate $N$, the number of monomers in projectile fragments $M$, and the total number of projectiles used to grow the aggregate $C$. These parameters are related by the simple equation $N=CM$.

Thus, we have carried out three main studies. In each study, we keep one parameter constant, vary another over a specified range, and set the third such that $N=CM$. The first study sets $N=300$ and varies $M=$ 1, 3, 5, 10, 15, 20, 30, 50, 60, 75, 100, 150, 300 (or equivalently $C=$ 300, 100, 60, 30, 20, 15, 10, 6, 5, 4, 3, 2, 1). This set will be referred to as constant aggregate size (CAS) and is shown in green throughout this paper. The second study sets $C=30$ and varies $M=$ 1, 3, 5, 10, 15, 20, 30, 50, 60 (or equivalently $N=$ 30, 90, 150, 300, 450, 600, 900, 1500, 1800). This set will be referred to as constant number of projectiles (CNP), and is shown in blue. Lastly, we set $M=20$ and vary $C=$ 1, 2, 3, 4, 5, 6, 10, 15, 20, 30 (or equivalently $N=$ 20, 40, 60, 80, 100, 120, 200, 300, 400, 600). This set will be referred to as constant fragment size (CFS) and is shown in pink.

In addition to these sets, we ran an additional CAS set for $M=$ 3, 15, and 100 where fragments are treated as rigid bodies instead of loosely bound aggregates to study how much fragments restructure upon impact. This set will be referred to as welded constant aggregate size (wCAS) and is shown in yellow. 

Each $(N,M,C)$ combination was simulated 30 times with a different random seed (i.e., different projectile fragments, fragment velocities, directions, and collision offsets). The presented results are averaged over those 30 runs. Error bars represent the standard error.

\subsection{Quantifying Aggregate Structure}
\label{subsec:structureQuantification}

The structures of the final aggregates are quantified using several metrics previously applied and evaluated by \citet{Kolanz2026}, including the equivalent ellipsoid porosity $\mathcal{P}_{abc}$, the gyration radius porosity $\mathcal{P}_{KBM}$, the geometric-cross-section-based porosity $\mathcal{P}_{gcs}$, and the fractal dimension $\mathcal{D}_{f}$. 
In addition to these metrics, the asymmetry parameter $\mathcal{A}$ \citep{Draine2024b} and stretch parameter $\mathcal{S}$ \citep{Draine2024a} are also calculated. 

\citet{Draine2024b} and \citet{Shen2008} define the dimensionless quantities

\begin{equation}
    \alpha _i \equiv \frac{I_i}{0.4m_{agg}r^{2}_{eff}}
\end{equation}

\noindent where $\mathbf{I}$ is the moment of inertia tensor of the aggregate with eigenvalues $I_1 \ge I_2 \ge I_3$, $r_{eff}$ is the radius of the single equivalent sphere and is defined as $r_{eff} \equiv [3m_{agg}/({4\pi\rho}) ]^{1/3}$, and $m_{agg}$ is the total mass of the aggregate. Using these, the asymmetry parameter is 

\begin{equation}
    \mathcal{A} \equiv \left( \frac{\alpha_1}{\alpha_2 + \alpha_3 - \alpha_1} \right)^{1/2}.
\end{equation}

The asymmetry parameter is at a minimum of $\mathcal{A}=1$ for a sphere and $\mathcal{A}>1$ for elongated or flattened shapes. 

In \citet{Draine2024b} the asymmetry parameter and porosity of interstellar dust are constrained using starlight polarization and polarized thermal emission. Draine finds that observed polarization can only be explained if the asymmetry and porosity (they used $\mathcal{P}_{abc}$ specifically) of interstellar dust follow the relationship

\begin{equation}
    \mathcal{A} \gtrsim 1+\frac{0.49}{(1-\mathcal{P}_{abc})^{4/3}}.
    \label{eq:AvsP}
\end{equation}

The stretch parameter is introduced as a way to distinguish flattened or oblate ($1/\sqrt{2} < \mathcal{S} < 1$) from elongated or prolate ($\mathcal{S}>1$) shapes. It is defined as

\begin{equation}
    \mathcal{S} \equiv \frac{\alpha_{2}}{(\alpha_{1}\alpha_{3})^{1/2}}.
\end{equation}

\section{Results}
\label{sec:results}

\begin{figure*}[!t]
    \centering
    \includegraphics[width=0.98\textwidth,keepaspectratio]{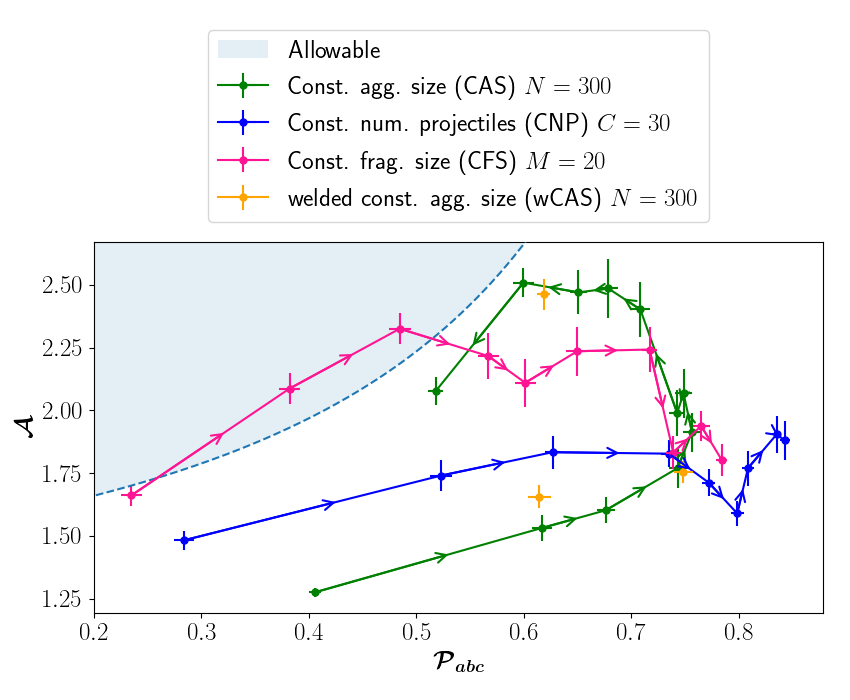}

    \caption{Asymmetry parameter vs equivalent ellipsoid porosity for all types of simulations. The green line represents CAS aggregates grown to a constant final size of $N=300$ monomers, parameterized over the fragment sizes $M=$ 1, 3, 5, 10, 15, 20, 30, 50, 60, 75, 100, 150, and 300. The blue line represents CNP aggregates grown with $C=30$ projectiles, parameterized over the fragment sizes $M=$ 1, 3, 5, 10, 15, 20, 30, 50, and 60. The pink line represents CFS aggregates grown with a constant fragment size of $M=20$ monomers, parameterized over the number of projectiles $C=$ 1, 2, 3, 4, 5, 6, 10, 15, 20, and 30. The yellow points represent wCAS aggregates, which are the same as the green but each projectile is treated as a rigid body, for projectiles of $M=$ 3, 15, and 100 monomers.}
 
        \label{fig:AvsP}
\end{figure*}

We find that the structure of an aggregate grown with sequential collisions is highly dependent on the structure and shape of its constituent projectiles. As seen in Figures~\ref{fig:CASPabcVsM}, \ref{fig:CNPPabcVsM}, and \ref{fig:PabcVsN}, $\mathcal{P}_{abc}$ doubles over two orders of magnitude in projectile size, an impressive difference for a parameter that ranges from $0 \le \mathcal{P}_{abc} \le 1$. Thus, projectile shape is a significantly greater driver of porosity and geometry than temperature. \citet{Kolanz2026} find that $\mathcal{P}_{abc}$ varies only by 0.05 ($\sim10\%$ of the variation seen in this study) over two orders of magnitude in temperature. 

The main difference between this work and \citet{Kolanz2026} is that the present study uses non-spherical projectiles. For $M>1$, projectiles are non-spherical, and thus cannot roll as easily into existing voids in the target aggregate. For $M=1$, projectiles are perfectly spherical monomers. Being perfect spheres, these projectiles can roll after their initial impact and more easily find and fill existing voids in an aggregate. All the aggregates discussed in \citet{Kolanz2026} were grown with $M=1$, and all $M=1$ aggregates in this study come from \citet{Kolanz2026}.

Additionally, we find that treating aggregates as rigid bodies rather than loosely bound aggregates has little effect on the final aggregate structure for the thermal velocities considered in this study. The slight differences seen are not statistically significant and appear to be dominated by the chaotic nature of the N-body problem.

\begin{figure*}
    \centering
    \includegraphics[width=\textwidth,height=0.93\textheight,keepaspectratio]{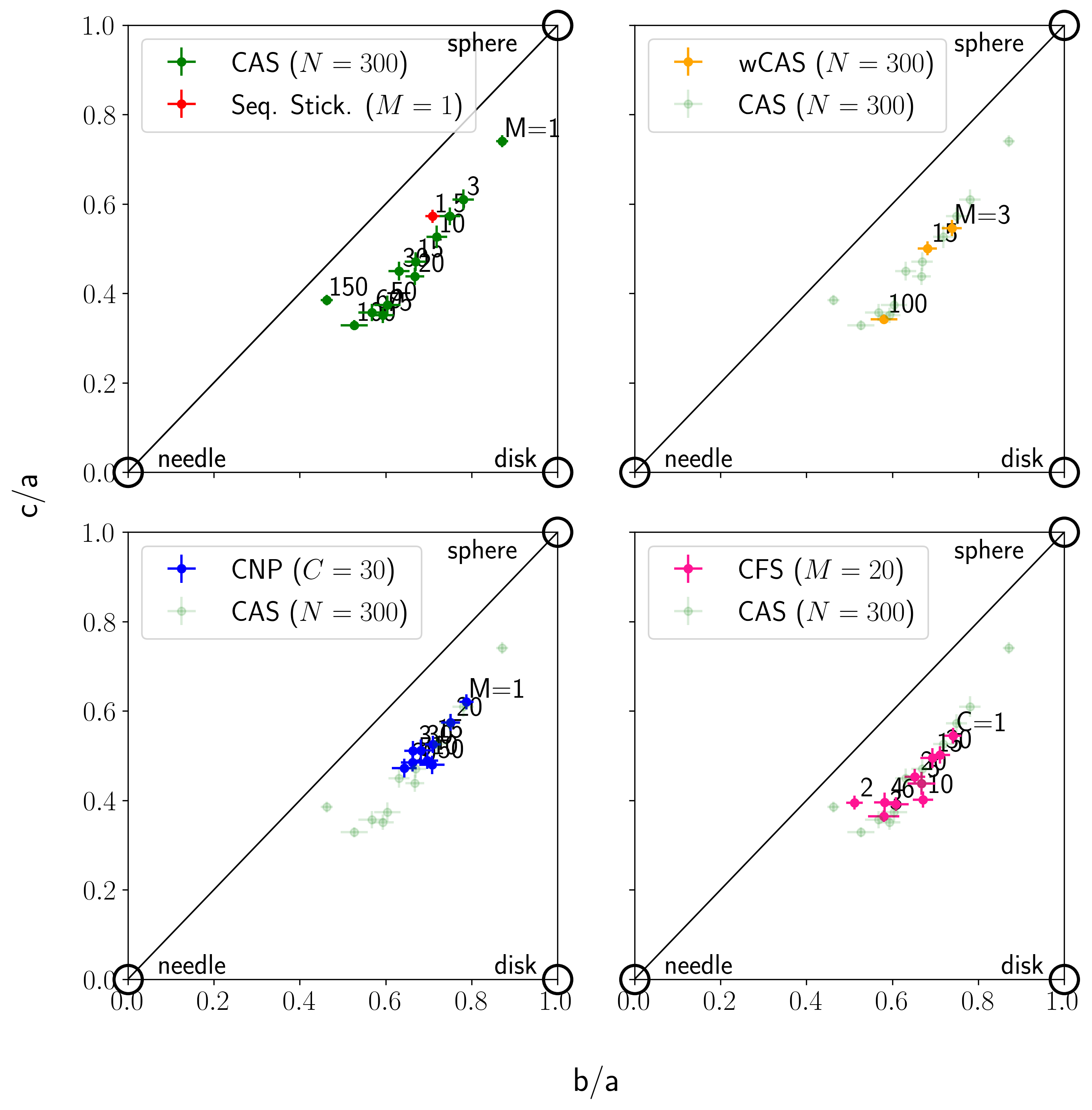}

    \caption{Ratios of the equivalent ellipsoid semi-major axes, $c/a$ vs $b/a$ for all aggregates in this study, where $a>b>c$. Given this constraint, the area above $y=x$ is forbidden. The closer an aggregate is to $(0,0)$, the closer it is to a one-dimensional needle. The closer it is to $(1,0)$, the closer it is to a plane. The closer it is to $(1,1)$, the closer it is to a sphere. Numbers next to the points indicate the projectile size $M$, except in the bottom-right panel, where they indicate the number of projectiles $C$. CAS aggregates with $N=300$ are plotted in all panels for comparison. Note that for CAS aggregates, $M=1$ and $M=300$ are the same point.} 
        \label{fig:axes_ratios}
\end{figure*}

In the following, we analyze in greater detail the effects of projectile size ($M$), number of projectiles ($C$), and final aggregate size ($N$) on the structure of the aggregates. To quantify structure, we use four different porosity metrics defined in \citet{Kolanz2026}, as well as the two shape parameters defined in Section~\ref{subsec:structureQuantification}.

\subsection{Constant Aggregate Size (CAS)}
\label{subsec:ConstFinalSize}

When grown to a constant final size of $N=300$ monomers, the structure of our CAS aggregates is cyclic over fragment size $M$ and number of projectiles $C$. This is shown for $\mathcal{P}_{abc}$ in Figure~\ref{fig:CASPabcVsM}, which is plotted with images of the aggregates to better visualize the cyclic nature of their structure. It is also shown for all metrics in the green lines of Figures~\ref{fig:structural_metrics_vs_M} and \ref{fig:structural_metrics_vs_C}. 

The cyclic behavior is not surprising, as an ($N=300$, $M=1$) aggregate is identical to an ($N=300$, $M=300$) aggregate grown at $1000$ K in \citet{Kolanz2026}. The rest of the variation in intermediate $M$ values is due to differences in non-spherical projectiles. Starting at $M=1$, projectile monomers are able to roll into existing gaps until multiple contacts are made between the new monomer and the aggregate. At $M=3$, projectiles start with contacts, and thus cannot make as many additional contacts upon impact with the target aggregate. As $M$ gets larger, even fewer new contacts per existing contact and projectile are made with the target aggregate, leading to a higher porosity. Eventually this effect is balanced out and overcome as $M$ gets even larger (and $C$ gets smaller). At this point, the limited number of total projectiles in the aggregate means there are less porous configurations possible, and so porosity goes back down.

Since porosity and fractal dimension are anticorrelated (a higher porosity corresponds to a lower fractal dimension and vice versa), it is unsurprising that the fractal dimension is also cyclic, but with a minimum instead of a maximum. This anticorrelation is seen in all plots of $\mathcal{P}$ and $\mathcal{D}_f$, and so further discussion will focus on the porosity, but is still applicable to the fractal dimension.

A similar cyclic trend is seen with $\mathcal{A}$ and $\mathcal{S}$ for CAS aggregates in Figure~\ref{fig:structural_metrics_vs_M}. However, the peak occurs much later, around $M\sim(60-100)$. This difference reflects how $\mathcal{A}$ and $\mathcal{S}$ are dependent only on the geometry of the aggregate and not the internal structure such as $\mathcal{P}_{abc}$, $\mathcal{P}_{KBM}$, $\mathcal{P}_{gcs}$, and $\mathcal{D}_{f}$. Metrics that focus on the geometry are not sensitive to the different internal voids caused by fragments of different sizes. Thus, void size peaks for $M\sim15$, but the fewer projectiles in an aggregate, the more aspherical its final structure. This tells us an aggregate's geometry is highly dependent on the number of projectiles.

\Needspace{4\baselineskip}

\subsection{Constant Number of Projectiles (CNP)}
\label{subsec:ConstNumberProjectiles}

Due to the dependence of an aggregate's final structure on the number of projectiles, we also study aggregates grown with a constant number of projectiles $C=30$. These CNP aggregates and their $\mathcal{P}_{abc}$ values are shown in Figure~\ref{fig:CNPPabcVsM}. All structural metrics for the CNP aggregates are shown by the blue lines of Figures~\ref{fig:structural_metrics_vs_M} and \ref{fig:structural_metrics_vs_N}. 

The porosity of these aggregates grows monotonically with $M$ and $N$, as does their overall size, visualized in Figure~\ref{fig:CNPPabcVsM}. However, porosity and size are not simply correlated, highlighting the growth path in the structure of the final aggregate. If we compare the $M=1$ points in Figure~\ref{fig:structural_metrics_vs_M}a, we see a difference in $\mathcal{P}_{abc}$ of around 0.1 when aggregates are grown with 30 vs.\ 300 monomers. If we instead compare aggregates grown with 30 projectiles of 10 monomers each and those grown with 300 individual monomers, the porosity differs by around 0.35. The additional difference is due to larger voids in aggregates made up of fragments, as fragments tend not to roll and restructure upon impact as a monomer would.

The observation in the previous section that an aggregate's geometry is highly dependent on the number of projectiles $C$ rather than fragment size $M$ is consistent here, where $\mathcal{A}$ and $\mathcal{S}$ are mostly constant for a constant number of non-spherical projectiles of all sizes.

\subsection{Constant Fragment Size (CFS)}
\label{subsec:ConstFragSize}

Additionally, we grew aggregates with various numbers of projectiles made up of $M=20$ monomer fragments. These CFS aggregates, along with their porosities, are shown in Figure~\ref{fig:PabcVsN}. The dependence of all structural metrics on $C$ and $N$ for the CFS aggregates is shown in the pink lines of Figures~\ref{fig:structural_metrics_vs_C} and \ref{fig:structural_metrics_vs_N}. 

Porosity increases monotonically with $C$ and $N$ for CFS aggregates, consistent with the porosity of CNP aggregates.   

On the other hand, the asymmetry and stretch parameters now follow the CAS aggregates. This is visualized in the bottom panels of Figure~\ref{fig:structural_metrics_vs_C}, and again indicates these parameters (and thus an aggregate's geometry) are highly dependent on the number of projectiles $C$.

\subsection{Aggregate Shape}
\label{subsec:AggGeometry}

None of our aggregates with $N>60$ are within the asymmetry-porosity space allowed by Equation~\ref{eq:AvsP}, shown in Figure~\ref{fig:AvsP}. Only the $N=40$ and $N=60$ CFS aggregates are within the region. The closest aggregates to the allowable region are the ones formed by three or fewer fragments, which seems to be a sweet spot for asymmetry vs porosity. With only three projectiles, the final aggregate has a good chance of those projectile fragments connecting along a single axis. With two projectiles, even though they must connect along a single axis, the final aggregate is less asymmetric than those with three projectiles. With four projectiles, the final aggregate is unlikely to form along a single axis and thus, on average, does not form as asymmetric a structure as with two or three projectiles. 

This is consistent with the results of \citet{Draine2024b}, who found three-sphere configurations have polarization properties closer to observational constraints than two spheres.

An intuitive way to discuss the geometry of aggregates is in ellipsoid axis space. Figure~\ref{fig:axes_ratios} shows $c/a$ vs $b/a$, where $a>b>c$ are the axes of the equivalent ellipsoid fit to each aggregate. In this space, all shapes fall below the $c/a=b/a$ line by definition. Locations near $(b/a,c/a)=(0,0)$, $(1,0)$, and $(1,1)$ correspond to a needle, a flat disk, and a sphere, respectively.

In general, for CAS aggregates grown to $N=300$ particles, a smaller fragment size ($M$) and more projectiles ($C$) produce more spherical final aggregates. Aggregates then become more needle-like as we increase $M$ and decrease $C$. These aggregates have a similar distance from the disk corner, except for a fragment size of $M=150$, which is less disk-like than the rest. 

CNP aggregates grown with a constant number of projectiles $C=30$ tend to cluster in the same area regardless of the fragment size ($M$) used. 

CFS aggregates grown with $M=20$ monomer fragments follow the same trend as CAS aggregates, where more non-spherical projectiles $C>1$ produce more spherical aggregates, and fewer projectiles produce more needle-like aggregates. Again, the aggregates formed with two projectiles are less disk-like than the rest of the aggregates.

Thus, differences in axis space seem to primarily be due to the number of projectiles $C$ making up the final aggregate, as with the asymmetry and stretch parameters. 

As in previous sections, the rigid body wCAS aggregates are consistent with the bound CAS aggregates.

\section{Summary and Discussion} \label{sec:summary}

We have numerically grown aggregates using sequential collisions to study the effect of non-spherical projectiles on aggregate structure. We find that aggregates grown with non-spherical projectiles have significantly different structures than those grown with spherical projectiles. 

At thermal velocities, spherical projectiles are able to roll on the target aggregate and fill voids, while non-spherical projectiles cannot roll, and do not have enough energy to restructure upon impact, mostly sticking where they impacted. Thus, aggregates grown with non-spherical projectiles are less dense (higher porosity, lower fractal dimension) than those grown with spherical projectiles. We see a similar effect on the asymmetry and stretch parameters. Non-spherical projectiles can stick where they land, leading to relatively less spherical final aggregates.

We also find that an aggregate's final structure is correlated with three variables: the number of projectiles that make up the aggregate $C$, the size of projectiles $M$, and the final size of the aggregate $N$. 

The porosity is positively correlated with both the size of projectiles $M$ and number of projectiles $C$ in our CNP and CFS aggregates. This is seen in the blue lines of Figures~\ref{fig:structural_metrics_vs_M} and \ref{fig:structural_metrics_vs_N} and in the pink lines of Figures~\ref{fig:structural_metrics_vs_C} and~\ref{fig:structural_metrics_vs_N}. This is due to larger projectiles and more projectiles both being able to create larger voids in the final aggregate. 

The correlation of the asymmetry and stretch parameters with the size of projectiles $M$ and number of projectiles $C$ is more complicated. The asymmetry and stretch parameters depend only weakly on projectile size $M$. On the other hand, varying the number of projectiles $C$ has a non-linear effect on these parameters. In this case, asymmetry and stretch parameters peak for $1<C \lesssim 10$. At low $C$, aggregates formed from only a few projectiles can adopt only a limited number of highly asymmetric configurations. As $C$ increases beyond approximately ten, the number of possible configurations increases, and the average aggregate becomes more spherical. 

For CAS aggregates grown to a final size of $N=300$ (green lines of Figures~\ref{fig:structural_metrics_vs_M} and~\ref{fig:structural_metrics_vs_C}), porosities peak for $M$ on the order of ten. This peak is caused by the competing effects of projectile size $M$ and number of projectiles $C$. Increasing both $M$ and $C$ increases porosity, but to keep $N=300$ and satisfy $N=CM$, one of $M$ or $C$ must decrease as the other increases.

Fragments treated as rigid bodies (wCAS aggregates) do not form aggregates appreciably different from those formed by fragments held together solely by attractive forces when collision velocities are thermal. This is seen in the yellow dots of Figures~\ref{fig:structural_metrics_vs_M},~\ref{fig:AvsP}, and~\ref{fig:axes_ratios}, which agree with their CAS counterparts.

Overall, the geometry of aggregates (asymmetry parameter, stretch parameter, ellipsoid axis space) depends heavily on the number of projectiles $C$, with more projectiles corresponding to a more spherical final aggregate. The geometry of aggregates is only weakly tied to the size of fragments $M$. On the other hand, the porosity and fractal dimensions depend heavily on both the size of fragments $M$ and the number of projectiles $C$.

None of our final aggregates with $N>60$ monomers have structures that conform to the constraint (Equation~\eqref{eq:AvsP}) \citet{Draine2024b} found to explain starlight polarization and polarization from thermal emission. In agreement with \citet{Draine2024b}, aggregates formed with fewer fragments are closest to aligning with Equation~\eqref{eq:AvsP}, as they are less porous and more asymmetric. 

Again in agreement with \citet{Draine2024b}, our results suggest aggregates formed through coagulation need additional processing before they can be responsible for observed polarization. This is consistent with decades of results suggesting that dust is continually processed from the time it is formed in asymptotic giant branch stars \citep{Sandin2004,Gobrecht2016}, supernovae \citep{Slavin2015, Kirchschlager2024}, and molecular clouds \citep{Ossenkopf1993, Bergin1995, VallucciGoy2024} to when it is present in the interstellar medium \citep{Draine1990,Galliano2018}. Processes that could push aggregates toward being more asymmetric and less porous could include photolytic densification or grain-grain collisions as suggested by \citet{Draine2024b}, or processing in shocks. 

In the case of grain-grain collisions, aggregates could become flatter and less porous through superthermal collisions. The fragments themselves could also have different structures if the assumption of thermal equilibrium is relaxed during their growth, though more simulations are necessary to reach these conclusions. 

An additional source of uncertainty in this study is our interaction model. Although our model is relatively simple and inexpensive computationally, it has a smaller barrier to rolling than other models such as the Johnson-Kendall-Roberts (JKR) contact model \citep{Johnson1971}. A comparison of aggregate structures produced by sequential collisions using our model and the JKR model will be explored in future work. Still, having a smaller barrier to rolling means less porous final aggregates, so it is unlikely the JKR model will create aggregates that better align with Equation~\eqref{eq:AvsP}.

It is also unlikely that aggregates in nature grow through collisions with consistently sized projectiles, as has been done in this study. In reality, the number of monomers in a projectile will vary between collisions and change over time as the populations of smaller projectiles are used up. This will be addressed in future work (Kolanz and Lazzati in preparation).

\begin{acknowledgements}
LK and DL acknowledge Job Guidos' contributions to the development of DECCO and useful discussions during the preparation of this work. LK acknowledges financial support from the ARCS Foundation Oregon, as well as access to facilities and help developing the DECCO code from Dr. Doru Thom Popovici and Dr. Mauro Del Ben with the Applied Computing for Scientific Discovery group at LBNL. 
\end{acknowledgements}

%%%%%%%%%%%%%%%%%%%%%%%%%%%%%%%%%%%%%%%%%%%%%%%%%%
\bibliography{main}

@article{Kartiwa2023,
	title = {Review of Quaternion Differential Equations: Historical Development, Applications, and Future Direction},
	volume = {12},
	rights = {http://creativecommons.org/licenses/by/3.0/},
	issn = {2075-1680},
	url = {https://www.mdpi.com/2075-1680/12/5/483},
	doi = {10.3390/axioms12050483},
	shorttitle = {Review of Quaternion Differential Equations},
	pages = {483},
	number = {5},
	journal = {Axioms},
	publisher = {Multidisciplinary Digital Publishing Institute},
	author = {Kartiwa, Alit and Supriatna, Asep K. and Rusyaman, Endang and Sulaiman, Jumat},
	urldate = {2026-05-14},
	date = {2023-05},
	year = {2023},
	langid = {english},
}

@article{Draine2024b,
	title = {Sensitivity of Polarization to Grain Shape. {II}. Aggregates},
	volume = {969},
	issn = {0004-637X},
	url = {https://ui.adsabs.harvard.edu/abs/2024ApJ...969...92D},
	doi = {10.3847/1538-4357/ad3b9a},
	pages = {92},
	journaltitle = {The Astrophysical Journal},
	journal = {The Astrophysical Journal},
	publisher = {{IOP}},
	author = {Draine, B. T.},
	urldate = {2025-11-05},
	date = {2024-07-01},
	year = {2024},
	note = {{ADS} Bibcode: 2024ApJ...969...92D},
}

@article{Draine2024a,
	title = {Sensitivity of Polarization to Grain Shape. I. Convex Shapes},
	volume = {961},
	issn = {0004-637X},
	url = {https://doi.org/10.3847/1538-4357/ad0463},
	doi = {10.3847/1538-4357/ad0463},
	pages = {103},
	number = {1},
	journaltitle = {The Astrophysical Journal},
	journal = {The Astrophysical Journal},
	shortjournal = {{ApJ}},
	publisher = {The American Astronomical Society},
	author = {Draine, B. T.},
	urldate = {2026-05-18},
	date = {2024-01},
	year = {2024},
	langid = {english},
}

@article{Kolanz2026,
	title = {Effect of temperature on the structure of porous dust aggregates formed by coagulation},
	volume = {9},
	url = {https://astro.theoj.org/article/158771-effect-of-temperature-on-the-structure-of-porous-dust-aggregates-formed-by-coagulation},
	doi = {10.33232/001c.158771},
	journal = {The Open Journal of Astrophysics},
	shortjournal = {The Open Journal of Astrophysics},
	publisher = {Maynooth Academic Publishing},
	author = {Kolanz, Lucas and Lazzati, Davide and Guidos, Job},
	urldate = {2026-04-29},
	date = {2026-03-09},
	year = {2026},
	langid = {english},
}

@article{Johnson1971,
   author = {K.L. Johnson and K Kendall and A.D. Roberts},
   doi = {10.1098/rspa.1971.0141},
   issn = {0080-4630},
   issue = {1558},
   journal = {Proceedings of the Royal Society of London. A. Mathematical and Physical Sciences},
   pages = {301-313},
   title = {Surface energy and the contact of elastic solids},
   volume = {324},
   year = {1971},
}

@article{Brownlee1985,
   author = {D E Brownlee},
   journal = {Annual Review of Earth and Planetary Sciences},
   pages = {147-173},
   title = {COSMIC DUST : COLLECTION AND RESEARCH},
   volume = {13},
   url = {https://www.annualreviews.org/doi/pdf/10.1146/annurev.ea.13.050185.001051},
   year = {1985},
}

@article{Wurm2000,
   author = {Gerhard Wurm and Jürgen Blum},
   doi = {10.1086/312447},
   issn = {0004637X},
   issue = {1},
   journal = {The Astrophysical Journal},
   pages = {L57-L60},
   title = {An Experimental Study on the Structure of Cosmic Dust Aggregates and Their Alignment by Motion Relative to Gas},
   volume = {529},
   year = {2000},
}

@article{Wada2007,
   author = {Koji Wada and Hidekazu Tanaka and Toru Suyama and Hiroshi Kimura and Tetsuo Yamamoto},
   doi = {10.1086/514332},
   issn = {0004-637X},
   issue = {1},
   journal = {The Astrophysical Journal},
   pages = {320-333},
   title = {Numerical Simulation of Dust Aggregate Collisions. I. Compression and Disruption of Two‐Dimensional Aggregates},
   volume = {661},
   year = {2007},
}

@article{Suyama2008,
   author = {Toru Suyama and Koji Wada and Hidekazu Tanaka},
   doi = {10.1086/590143},
   issn = {0004-637X},
   issue = {2},
   journal = {The Astrophysical Journal},
   pages = {1310-1322},
   title = {Numerical Simulation of Density Evolution of Dust Aggregates in Protoplanetary Disks. I. Head‐on Collisions},
   volume = {684},
   year = {2008},
}

@article{Shen2008,
   author = {Yue Shen and B. T. Draine and Eric T. Johnson},
   doi = {10.1086/592765},
   issn = {0004-637X},
   issue = {1},
   journal = {The Astrophysical Journal},
   pages = {260-275},
   title = {Modeling Porous Dust Grains with Ballistic Aggregates. I. Geometry and Optical Properties},
   volume = {689},
   year = {2008},
}

@article{Slavin2015,
   author = {Jonathan D. Slavin and Eli Dwek and Anthony P. Jones},
   doi = {10.1088/0004-637X/803/1/7},
   issn = {15384357},
   issue = {1},
   journal = {Astrophysical Journal},
   pages = {7},
   publisher = {IOP Publishing},
   title = {Destruction of interstellar dust in evolving supernova remnant shock waves},
   volume = {803},
   url = {http://dx.doi.org/10.1088/0004-637X/803/1/7},
   year = {2015},
}

@article{Ysard2018,
   author = {N. Ysard and A. P. Jones and K. Demyk and T. Boutéraon and M. Koehler},
   doi = {10.1051/0004-6361/201833386},
   issn = {14320746},
   journal = {Astronomy and Astrophysics},
   pages = {1-19},
   title = {The optical properties of dust: The effects of composition, size, and structure},
   volume = {617},
   year = {2018},
}

@article{kirchschlager2024,
	title = {From total destruction to complete survival: dust processing at different evolutionary stages in the supernova remnant Cassiopeia A},
	volume = {528},
	issn = {0035-8711},
	url = {https://doi.org/10.1093/mnras/stae365},
	doi = {10.1093/mnras/stae365},
	shorttitle = {From total destruction to complete survival},
	pages = {5364--5376},
	number = {3},
	journal = {Monthly Notices of the Royal Astronomical Society},
	shortjournal = {Monthly Notices of the Royal Astronomical Society},
	author = {Kirchschlager, Florian and Sartorio, Nina S and De Looze, Ilse and Barlow, M J and Schmidt, Franziska D and Priestley, Felix D},
	urldate = {2024-10-11},
	date = {2024-03-01},
        year = {2024},
}

@ARTICLE{Guidos2025,
       author = {{Guidos}, Job and {Kolanz}, Lucas and {Lazzati}, Davide},
        title = "{Discrete element simulations of self-gravitating rubble pile collisions: the effects of non-uniform particle size and rotation}",
      journal = {The Open Journal of Astrophysics},
         year = 2025,
        month = aug,
       volume = {8},
          eid = {119},
        pages = {119},
          doi = {10.33232/001c.143461},
archivePrefix = {arXiv},
       eprint = {2410.22189},
 primaryClass = {astro-ph.EP},
       adsurl = {https://ui.adsabs.harvard.edu/abs/2025OJAp....8E.119G}
}

@article{morstein2022,
	title = {Humidity-dependent lubrication of highly loaded contacts by graphite and a structural transition to turbostratic carbon},
	volume = {13},
	rights = {2022 The Author(s)},
	issn = {2041-1723},
	url = {https://www.nature.com/articles/s41467-022-33481-9},
	doi = {10.1038/s41467-022-33481-9},
	pages = {5958},
	number = {1},
	journal = {Nature Communications},
	shortjournal = {Nat Commun},
	author = {Morstein, Carina Elisabeth and Klemenz, Andreas and Dienwiebel, Martin and Moseler, Michael},
	urldate = {2025-08-29},
	date = {2022-10-10},
	year = {2022},
	langid = {english},
	note = {Publisher: Nature Publishing Group},
}

@article{Yan2004,
	title = {Dust Dynamics in Compressible Magnetohydrodynamic Turbulence},
	volume = {616},
	issn = {0004-637X},
	url = {https://iopscience.iop.org/article/10.1086/425111/meta},
	doi = {10.1086/425111},
	pages = {895},
	number = {2},
	journal = {The Astrophysical Journal},
	shortjournal = {{ApJ}},
	author = {Yan, Huirong and Lazarian, A. and Draine, B. T.},
	urldate = {2025-11-14},
	date = {2004-12-01},
	year = {2004},
	langid = {english},
	note = {Publisher: {IOP} Publishing},
}

@article{Gonzalez2025,
	title = {Dusty clump survival in supernova ejecta - Dust-mediated growth versus crushing by the reverse shock},
	volume = {702},
	rights = {© The Authors 2025},
	issn = {0004-6361, 1432-0746},
	url = {https://www.aanda.org/articles/aa/abs/2025/10/aa56389-25/aa56389-25.html},
	doi = {10.1051/0004-6361/202556389},
	pages = {L6},
	journal = {Astronomy \& Astrophysics},
	shortjournal = {A\&A},
	publisher = {{EDP} Sciences},
	author = {Martínez-González, Sergio},
	urldate = {2025-11-03},
	date = {2025-10-01},
	year = {2025},
	langid = {english},
}

@article{Dopcke2011,
	title = {{THE} {EFFECT} {OF} {DUST} {COOLING} {ON} {LOW}-{METALLICITY} {STAR}-{FORMING} {CLOUDS}},
	volume = {729},
	issn = {2041-8205},
	url = {https://doi.org/10.1088/2041-8205/729/1/L3},
	doi = {10.1088/2041-8205/729/1/L3},
	pages = {L3},
	number = {1},
	journal = {The Astrophysical Journal Letters},
	shortjournal = {{ApJL}},
	publisher = {The American Astronomical Society},
	author = {Dopcke, Gustavo and Glover, Simon C. O. and Clark, Paul C. and Klessen, Ralf S.},
	urldate = {2026-04-07},
	date = {2011-02},
	year = {2025},
	langid = {english},
}

@article{Govender2023,
	title = {The influence of cohesion on polyhedral shapes during mixing in a drum},
	volume = {270},
	issn = {0009-2509},
	url = {https://www.sciencedirect.com/science/article/pii/S0009250923000556},
	doi = {10.1016/j.ces.2023.118499},
	urldate = {2026-02-06},
	journal = {Chemical Engineering Science},
	author = {Govender, Nicolin and Kobyłka, Rafał and Khinast, Johannes},
	month = apr,
	year = {2023},
	pages = {118499},
}

@article{Hou2023,
	title = {Discrete {Element} {Analysis} of {Shape} {Effect} on the {Shear} {Behaviors} of {Ballast}},
	volume = {13},
	issn = {2045-2322},
	url = {https://pmc.ncbi.nlm.nih.gov/articles/PMC10491666/},
	doi = {10.1038/s41598-023-42070-9},
	urldate = {2026-02-06},
	journal = {Scientific Reports},
	author = {Hou, Wenjie and Li, Ang and Song, Weimin},
	month = sep,
	year = {2023},
	pmid = {37684325},
	pmcid = {PMC10491666},
	pages = {14810},
}

@article{Tangri2019,
	title = {Hopper discharge of elongated particles of varying aspect ratio: {Experiments} and {DEM} simulations},
	volume = {4},
	issn = {2590-1400},
	shorttitle = {Hopper discharge of elongated particles of varying aspect ratio},
	url = {https://www.sciencedirect.com/science/article/pii/S2590140019300474},
	doi = {10.1016/j.cesx.2019.100040},
	urldate = {2026-04-07},
	journal = {Chemical Engineering Science: X},
	author = {Tangri, Henna and Guo, Yu and Curtis, Jennifer S.},
	month = nov,
	year = {2019},
	pages = {100040},
}

@article{Verlet1967,
  title = {Computer "Experiments" on Classical Fluids. I. Thermodynamical Properties of Lennard-Jones Molecules},
  author = {Verlet, Loup},
  journal = {Phys. Rev.},
  volume = {159},
  issue = {1},
  pages = {98--103},
  numpages = {0},
  year = {1967},
  month = {Jul},
  publisher = {American Physical Society},
  doi = {10.1103/PhysRev.159.98},
  url = {https://link.aps.org/doi/10.1103/PhysRev.159.98}
}

@article{Favier1999,
	title = {Shape representation of axi-symmetrical, non-spherical particles in discrete element simulation using multi-element model particles},
	volume = {16},
	doi = {10.1108/02644409910271894},
	journal = {Engineering Computations},
	author = {Favier, John and Abbaspour-Fard, Mohammad and Kremmer, M. and Raji, A.O.},
	month = jun,
	year = {1999},
	pages = {467--480},
}

@article{Bergin1995,
	title = {Gas-{Phase} {Chemistry} in {Dense} {Interstellar} {Clouds} {Including} {Grain} {Surface} {Molecular} {Depletion} and {Desorption}},
	volume = {441},
	issn = {0004-637X},
	url = {https://ui.adsabs.harvard.edu/abs/1995ApJ...441..222B},
	doi = {10.1086/175351},
	urldate = {2024-11-13},
	journal = {The Astrophysical Journal},
	publisher = {IOP},
	author = {Bergin, E. A. and Langer, W. D. and Goldsmith, P. F.},
	month = mar,
	year = {1995},
	note = {ADS Bibcode: 1995ApJ...441..222B},
	pages = {222},
}

@article{Ossenkopf1993,
	title = {Dust coagulation in dense molecular clouds : the formation of fluffy aggregates.},
	volume = {280},
	issn = {0004-6361},
	shorttitle = {Dust coagulation in dense molecular clouds},
	url = {https://ui.adsabs.harvard.edu/abs/1993A&A...280..617O},
	urldate = {2025-11-30},
	journal = {Astronomy and Astrophysics},
	publisher = {EDP},
	author = {Ossenkopf, V.},
	month = dec,
	year = {1993},
	note = {ADS Bibcode: 1993A\&A...280..617O},
	pages = {617--646},
}

@article{Sandin2004,
	title = {Three-component modeling of {C}-rich {AGB} star winds - {III}. {Micro}-physics of drift-dependent dust formation},
	volume = {413},
	copyright = {© ESO, 2004},
	issn = {0004-6361, 1432-0746},
	url = {https://www.aanda.org/articles/aa/abs/2004/03/aah4438/aah4438.html},
	doi = {10.1051/0004-6361:20031530},
	language = {en},
	number = {3},
	urldate = {2026-07-15},
	journal = {Astronomy \& Astrophysics},
	publisher = {EDP Sciences},
	author = {Sandin, C. and Höfner, S.},
	month = jan,
	year = {2004},
	pages = {789--798},
}

@article{Gobrecht2016,
	title = {Dust formation in the oxygen-rich {AGB} star {IK} {Tauri}},
	volume = {585},
	issn = {0004-6361},
	url = {https://ui.adsabs.harvard.edu/abs/2016A&A...585A...6G},
	doi = {10.1051/0004-6361/201425363},
	urldate = {2026-07-15},
	journal = {Astronomy and Astrophysics},
	publisher = {EDP},
	author = {Gobrecht, D. and Cherchneff, I. and Sarangi, A. and Plane, J. M. C. and Bromley, S. T.},
	month = jan,
	year = {2016},
	note = {ADS Bibcode: 2016A\&A...585A...6G},
	pages = {A6},
}

@inproceedings{Draine1990,
    author    = {Draine, Bruce T.},
    title     = {Evolution of Interstellar Dust},
    booktitle = {The Evolution of the Interstellar Medium},
    editor    = {Blitz, Leo},
    series    = {Astronomical Society of the Pacific Conference Series},
    volume    = {12},
    pages     = {193--205},
    publisher = {Astronomical Society of the Pacific},
    address   = {San Francisco},
    year      = {1990}
}

@article{VallucciGoy2024,
	title = {Dust evolution during a protostellar collapse: {Influence} on the coupling between the neutral gas and magnetic field},
	volume = {690},
	copyright = {© The Authors 2024},
	issn = {0004-6361, 1432-0746},
	shorttitle = {Dust evolution during a protostellar collapse},
	url = {https://www.aanda.org/articles/aa/abs/2024/10/aa48268-23/aa48268-23.html},
	doi = {10.1051/0004-6361/202348268},
	language = {en},
	urldate = {2026-07-15},
	journal = {Astronomy \& Astrophysics},
	publisher = {EDP Sciences},
	author = {Vallucci-Goy, V. and Lebreuilly, U. and Hennebelle, P.},
	month = oct,
	year = {2024},
	pages = {A23},
}

@article{Galliano2018,
	title = {The {Interstellar} {Dust} {Properties} of {Nearby} {Galaxies}},
	volume = {56},
	issn = {0066-4146},
	url = {https://ui.adsabs.harvard.edu/abs/2018ARA&A..56..673G},
	doi = {10.1146/annurev-astro-081817-051900},
	urldate = {2026-07-15},
	journal = {Annual Review of Astronomy and Astrophysics},
	author = {Galliano, Frédéric and Galametz, Maud and Jones, Anthony P.},
	month = sep,
	year = {2018},
	note = {ADS Bibcode: 2018ARA\&A..56..673G},
	pages = {673--713},
}
% \input{main.bll}

%%%%%%%%%%%%%%%%%%%%%%%%%%%%%%%%%%%%%%%%%%%%%%%%%%

\end{document}